\documentclass[aps,preprint]{revtex4}
\usepackage{amssymb}
\usepackage{amsmath}

\input{tcilatex}

\begin{document}

\title{Thermomagnetic instability of standing flux-antiflux front in layered
type-II superconductors.}
\author{E. E. Dvash$^{1}$, I. Shapiro$^{1}$, B. Rosenstein$^{2}$ and B. Ya.
Shapiro$^{1}$}
\affiliation{$^{1}$Department of Physics, Institute of Superconductivity, Bar-Ilan
University, Ramat-Gan 52900, Israel}
\affiliation{$^{2}$Department of Electrophysics, National Chiao Tung University, Hsinchu,
Taiwan, R.O.C. }

\begin{abstract}
Stability of standing flux-antiflux front in anisotropic layered
superconductors is considered. We describe two assisting mechanisms
destabilizing the standing vortex-antivortex front. There are anisotropy of
the layered superconductors and the heat, released by the vortex dynamics.
We present the conditions of \ the front stability for various anisotropy
and heating parameters. We predict that even small vortex-antivortex heating
can result in front instability. The characteristic size of the unstable
pattern is estimated.
\end{abstract}

\maketitle

\section{Introduction}

Studies of patterns in the magnetic flux distribution in the type-II
superconductors are attracting the attention of many research groups \cite%
{Vlasko-Vlascov,Koblischka,Johansen,Duran} whose magneto-optical experiments
demonstrate that nonuniform flux penetration occurs. Patterns with
branch-like structures have been found in most of high $T_{c}$ materials,
like $YBa_{2}Cu_{3}O_{7-x}$\cite{Leiderer} and $Bi_{2}Sr_{2}CaCu_{2\
}O_{8+x} $ \cite{Yeshurun}. The nucleation of dendrite like patterns in $%
MgB_{2}$ films is another example \cite{Johansen02,Barkov}. These complex
structures consist of alternating low and high vortex density regions and
are found in a certain temperature window. Likewise, flux penetration in the
form of droplets separating areas of different densities of vortices has
been observed in $NbSe_{2}$ \cite{Marchevsky}. Usually the occurrence of
flux patterns in interfacial growth phenomena can be attributed to a
diffusion driven, long-wavelength instability of a straight front, similar
to the Mullens-Sekerka instability \cite{Mullins} found in crystal growth.
The nucleation of nonuniform patterns associated with the propagation of a
flux front into a flux-free sample has been attributed to such an
interfacial instability. This results from a thermomagnetic coupling \cite%
{Johansen02,Barkov,Mints,Aranson01,Rakhmanov04} where a higher temperature
leads to a higher vortex mobility, enhanced flux flow, and hence a larger
heat generation.

\textit{On the other hand, the situation when the vortices interact in a
superconductor with the flux of the opposite sign is less theoretically
studied}. This flux configurations arises for example when a DC bias current
creates vortices and antivortices on the opposite side of the
superconducting strip \cite{Aranson}. Another example which is now under
intensive investigation arises upon exposing the previously magnetized
sample to the magnetic field of an opposite direction. According to the
experimental data \cite{Vlasko-Vlascov,Koblischka,Frello,Indebom,Duran}, the
boundary between vortices and antivortices exhibits many features suggestive
of a long wavelength instability. The cause of the instability at the
boundary between fluxes of opposite sign is still being debated. In
particular, Fisher at el. \cite{Fisher,Rakhmanov} proposed a non
thermomagnetic mechanism of instability caused by an in-plane anisotropy of
the vortex mobility. This mechanism of instability was carefully
reinvestigate by the van Saarloos et el \cite{Van Saarloos}. They confirmed
the finding of Fisher et el. \cite{Fisher,Rakhmanov} that standing
vortex-antivortex fronts have an instability to a modulated state, while the
moving fronts were found to be stable for all anisotropies. In fact however,
the flux-antiflux instability was experimentally detected in a system with
small \cite{Duran} and moderate anisotropy \cite{Koblischka}. Several years
latter this model was improved by an additional assumption of a step shape
and anisotropy of the voltage current characteristics \cite{Rakhmanov} and
explained the experimental result in moderate anisotropic superconductor $%
YBa_{2}Cu_{3}O_{7-\delta }$ . Unfortunately this assumption cannot explain
the instability in pure isotropic systems like $Nb$ and $MgB_{2}$ . On the
other hand the thermomagnetic mechanism can also be responsible for
flux-antiflux instability when the flux-antiflux front is heated both by the
vortex (antivortex) dynamics and by the vortex-antivortex annihilation.

In this paper we report on the thermomagnetic theory of the flux-antiflux
instability in anisotropic layered superconductor.

\section{Model and Basic Equations}

We start with a model of two-component vortex gas \cite{Bryksin} spatially
homogeneous along the $z$ axes, which is valid for the experimentally
interesting situation of the low magnetic field when typical spacing between
vortices $a_{0}$ essentially exceeds vortex-vortex (antivortex) interaction
radius $\xi $, and the vortex velocity depends only on the edge screening
current that is assumed to be homogeneously distributed across the sample.
The vortex-vortex repulsion, in this case, keeping the number of vortices,
cannot play a significant role. One must take into account both
vortex-antivortex annihilation and heat release accompanying this process.
We should also take into consideration heat absorption by the sample lattice
in order to prevent the rise of unlimited temperature.

The vortex-antivortex annihilation obeys the well-known master equations of
the recombination theory \cite{Landau10,Bakhanova}%
\begin{equation}
\frac{\partial n_{+}}{\partial t}+\nabla \left( n_{+}\mathbf{v}_{+}\right)
+gn_{+}n_{-}=0,  \label{n1}
\end{equation}%
\begin{equation}
\frac{\partial n_{-}}{\partial t}+\nabla \left( n_{-}\mathbf{v}_{-}\right)
+gn_{+}n_{-}=0  \label{n2}
\end{equation}%
\begin{equation}
g=\xi v,\ \ v=\func{mod}\left( \mathbf{v}_{+}-\mathbf{v}_{-}\right)
\label{n3}
\end{equation}%
where $n_{+}$ and $n_{-}$ are the vortex and antivortex densities,
respectively, $g$ is the ratio of recombination for vortices and
antivortices, $\xi $ is the cross section of the annihilation, which is of
the order of the coherence length of the superconductor, and $\mathbf{v}%
_{\pm }$ are the opposite directed vortex-antivortex velocities, which in
the creep regime are strongly temperature dependent:%
\begin{equation*}
\func{mod}\left( \mathbf{v}_{\pm }\right) =v_{\pm }=v_{\pm FF}\exp \left( -%
\frac{U}{T}\right) .
\end{equation*}%
\begin{equation}
\mathbf{J=}\frac{c}{4\pi }\nabla \times \mathbf{B,}\text{ }\mathbf{B}%
=\varphi _{0}\left( n_{+}-n_{-}\right) ,\text{ }\mathbf{v}_{\pm FF}=\frac{%
\mathbf{J}\times \varphi _{0}}{\eta c}.  \label{n4}
\end{equation}

Here, $U$ is a temperature-dependent pinning potential, $\varphi _{0}$ is
the unit flux, $\mathbf{J}$ is the electric current, and $\eta $ is the
viscosity of the vortices, which in an anisotropic system is a tensor with
different in plane and across the plane tensor components $\eta _{\alpha
\beta }$.

(It should be noted that our master (recombination) equations account
numbers of the \textit{topological charges} (vortex/antivortex cores). The
number of the vortices and antivortices is changed when they meet each other
at the distance of the order of the coherence length $\xi $ , rather than
the penetration length of the magnetic field $\lambda $. For example, the
same equations describe vortex-antivortex annihilation both in the
superfluid He4 and in superconducting films where the magnetic field is
uniform).

We wish to investigate an anisotropic system where the vortices velocity is
not necessarily parallel or perpendicular to the layers of the
superconductor. Therefor, we shall take the vortices velocity in a general
form%
\begin{equation}
v_{\alpha }=-\gamma _{\alpha \beta }\left( \frac{\partial B}{\partial
x_{\beta }}\right) .  \label{velocity}
\end{equation}

$\alpha ,\beta =x,y$

Here $\gamma =\varphi _{0}/4\pi \eta $ when%
\begin{equation}
\gamma =\eta ^{-1}=\eta _{0}^{-1}\left( 
\begin{array}{cc}
\cos ^{2}\vartheta +\alpha \sin ^{2}\vartheta & \cos \vartheta \sin
\vartheta \left( 1-\alpha \right) \\ 
\cos \vartheta \sin \vartheta \left( 1-\alpha \right) & \alpha \cos
^{2}\vartheta +\sin ^{2}\vartheta%
\end{array}%
\right)  \label{viscosity}
\end{equation}%
is the inverse tensor of the vortex viscosity \cite{Van Saarloos}, $\alpha \ 
$ is the anisotropy parameter of the system $\left( 0<\alpha <1\right) $, $%
\vartheta $ is the angle between the $x$ axis and the $a-b$ plain of the
layered structure (Fig. 1).

Assuming that the heat diffusion length exceeds the width of the slab \cite%
{Aranson01} one can complete the set of equations Eqs. (\ref{n1}) and (\ref%
{n2}) by the temperature transfer equation in the form%
\begin{equation}
\frac{\partial T}{\partial t}=\frac{\kappa _{T}}{C_{p}}\bigtriangleup T+%
\frac{1}{C_{p}}\frac{\delta Q}{\delta t}-\frac{\left( T-T_{0}\right) }{t_{R}}%
,  \label{n5}
\end{equation}%
where 
\begin{equation}
\frac{\partial Q}{\partial t}=W_{J}+W_{A}  \label{n6a}
\end{equation}%
\begin{equation}
W_{J}=\frac{\eta v^{2}}{2}\left( n_{+}+n_{-}\right) ;W_{A}=\xi v\frac{%
n_{+}n_{-}}{C_{p}}Q_{0}  \label{n6b}
\end{equation}%
is determined by the energy released both by vortex-antivortex dynamics $%
W_{J}$ and by vortex-antivortex annihilation $W_{A}$. Here $\kappa _{T}$ is
the heat conductivity, $C_{p}$ is the heat capacity, $T_{0}$ is the coolant
temperature, $Q_{0}$ is heat released by annihilation of a single
vortex-antivortex pair per unit vortex length, and $\ t_{R}$ is the
characteristic time of temperature relaxation.

The set of Eqs. (\ref{n1})-(\ref{n5}) completed by the boundary conditions
describes all features of the model.

\section{Spatial Distribution of Flux-antiflux Densities}

We consider the case of a restricted sample of length $D$. In this case, the
flux-antiflux interface in the stationary state is formed due to a balance
between flux-antiflux entering the sample from the opposite sides and their
annihilation in the middle point.

Introducing new variables%
\begin{equation*}
n_{+}/n_{m}=N_{+},\text{ }n_{-}/n_{m}=N_{-},\text{ }x/\Delta L\rightarrow x,
\end{equation*}%
\begin{equation}
\Delta L=\left( n_{m}\xi \right) ^{-1},\text{ }t/t_{0}=t^{\ast },\text{ }%
t_{0}=\frac{\eta \left( \Delta L\right) ^{2}}{n_{m}\varphi _{0}^{2}},
\label{n20}
\end{equation}%
\begin{equation}
b=N_{+}-N_{-},\text{ }N=N_{+}+N_{-},  \label{n23}
\end{equation}%
where $n_{m}$ is flux density at the interface point, and $\Delta L$ is the
characteristic width of the region in which the spatial distributions of
vortex and antivortex flux densities overlap, forming the interlayer where
the vortices of the opposite signs coexist. $\Delta L$ may be estimated as $%
\Delta L\simeq v/gn_{m}\sim \left( n_{m}\xi \right) ^{-1}$ (see Ref. \cite%
{Shapiro}), which is a microscopically large area where the total magnetic
induction is suppressed.

One obtain from Eqs. (\ref{n1})-(\ref{n2})%
\begin{equation}
\frac{\partial b}{\partial t^{\ast }}=\frac{\partial }{\partial x}\left( N%
\frac{\partial b}{\partial x}\right) ,  \label{n24}
\end{equation}%
\begin{equation}
\frac{\partial N}{\partial t^{\ast }}=\frac{\partial }{\partial x}\left( b%
\frac{\partial b}{\partial x}\right) -\left\vert N^{2}-b^{2}\right\vert
\left\vert \frac{\partial b}{\partial x}\right\vert ,  \label{n25}
\end{equation}%
where $b$ is the dimensionless magnetic induction.

In the stationary state we get for $N_{0}$ and $b_{0}$%
\begin{equation}
\frac{\partial }{\partial x}\left( N_{0}\frac{\partial b_{0}}{\partial x}%
\right) =0  \label{n24a}
\end{equation}%
\begin{equation}
\frac{\partial }{\partial x}\left( b_{0}\frac{\partial b_{0}}{\partial x}%
\right) -\left( N_{0}^{2}-b_{0}^{2}\right) \left\vert \frac{\partial b_{0}}{%
\partial x}\right\vert =0  \label{n24b}
\end{equation}%
\bigskip

Performing the integration in Eq. (\ref{n24a}) one obtain%
\begin{equation}
N_{0}\frac{\partial b_{0}}{\partial x}=-I.  \label{n27}
\end{equation}

here $I$ is a constant.

Substituting $N$ from Eq. (\ref{n27}) into Eq. (\ref{n24b}) and performing
the integration we immediately obtain the differential equation for $b$ in
the form 
\begin{equation}
-W+\frac{I}{2}\ln \left\vert \frac{I+W}{I-W}\right\vert =-\frac{b_{0}^{3}}{3}%
.  \label{n30}
\end{equation}%
where%
\begin{equation}
W=b_{0}\frac{\partial b_{0}}{\partial x}.  \label{n29}
\end{equation}

\subsection{Flux-antiflux Interface}

This equation may be solved analytically close to the interface line where $%
b_{0}$ goes to zero.

Assuming that the vortices and antivortices appear at the edges of the
samples separated by the distance $D$ (in dimensionless units) and assuming
the following boundary conditions%
\begin{equation*}
b_{0}\frac{\partial b_{0}}{\partial x}=-I,\text{ }N_{0-}=0\text{ at }x=-%
\frac{D}{2},
\end{equation*}%
\begin{equation}
b_{0}\frac{\partial b_{0}}{\partial x}=I,\text{ }N_{0+}=0\text{ at }x=\frac{D%
}{2},  \label{n31}
\end{equation}%
we obtain an asymptotically exact result for magnetic induction and vortices
density at the interface: Looking for the solution in the vicinity of the
flux-antiflux front $\left( X\rightarrow 0\right) $ in the form%
\begin{equation}
b_{0}\simeq a_{1}x+a_{2}x^{3};\text{ }N_{0}\simeq a_{3}+a_{4}x^{2}
\label{n14ab}
\end{equation}%
one obtains from Eqs. (\ref{n27})-(\ref{n29}) to the main order%
\begin{equation}
a_{1}=-I^{2/3},\text{ }a_{2}=\frac{I^{4/3}}{180},\text{ }a_{3}=I^{1/3},\text{
}a_{4}=\frac{I}{60}  \label{n14b}
\end{equation}%
It should be noted numerical simulation show that these formulas are valid
in a much wider region at front $\left( b_{0}=0\right) $ and can be
considered as an interpolation ones.

Assuming that the slope of the magnetic induction at the front $I\ll 1$ is
small one obtains for the characteristic size of the vortex-antivortex area $%
x_{c}$, where vortices and antivortices coexist $N_{0}\left( x_{c}\right)
=b_{0}\left( x_{c}\right) $ (see Eq.(\ref{n24a})) (see Fig.2). 
\begin{equation}
x_{c}=I^{-1/3}>>1  \label{interface}
\end{equation}

The dimensionless vortex velocity at the interface $u_{0}=I^{2/3}$ is a
constant. Returning to the dimension variables, we obtain for interface flux
velocity%
\begin{equation}
u_{\pm }\approx \frac{n_{m}^{2}\xi \varphi _{0}^{2}}{4\pi }\frac{\exp \left(
-U/T\right) }{\eta }I^{2/3},  \label{n33}
\end{equation}

\section{Overheating Instability}

Let us consider the stability of the vortex-antivortex interface with
respect to small deviations from its initial plane shape. For this we shall
take the Eqs. (\ref{n24}), (\ref{n25}), and (\ref{n5}) in the more general
form%
\begin{equation}
\frac{\partial b}{\partial t}-\nabla \left( N\mathbf{v}\right) =0
\label{n34}
\end{equation}%
\begin{equation}
\frac{\partial N}{\partial t}-\nabla \left( b\mathbf{v}\right) +\left(
N^{2}-b^{2}\right) \left\vert v\right\vert =0  \label{n35}
\end{equation}%
\begin{equation}
\frac{\partial \Theta }{\partial t}-\kappa \nabla ^{2}\Theta
-w_{A}-w_{J}+r\left( \Theta -1\right) =0  \label{n36}
\end{equation}%
where 
\begin{equation}
w_{A}=\left( N^{2}-b^{2}\right) S_{A}\left\vert v\right\vert
,w_{J}=NS_{J}v^{2}  \label{n36a}
\end{equation}
are the dimensionless annihilation and Joule heat terms$.$

Here $S_{A}\equiv \left( Q_{0}n_{m}/4\pi T_{0}C_{p}\right) ;S_{J}=$ $\varphi
_{0}^{2}n_{m}^{2}/16\pi ^{2}T_{0}C_{p}$ are the heating parameters,$%
Q_{0}\sim \varphi _{0}^{2}/\lambda ^{2}$ where $\lambda $ is the London
penetration length. The ratio $S_{J}/S_{A}\sim \lambda ^{2}n_{m}.$

It seems at first glance that the direct Joule term caused by vortex
(antivortex) motion always prevails. Really, for a sharp shape magnetic
induction front the vortex-antivortex annihilation term (overlapping) which
is proportional to the vortex (antivortex) density production $w_{A}\sim
vN_{+}N_{-}$ is small while vortex velocity $v\sim \nabla b$ is large.
Therefore the direct, Joule term which is proportional both to the sum of
the vortex and antivortex densities and to the square of the velocity $%
w_{J}=NS_{J}v^{2}$ significantly exceeds the annihilation term. \textit{%
However, in our case, when the slope of the magnetic induction profile is
small the annihilation term becomes essentially important.} In this case the
overlapping (annihilation) term is larger due to deep mutual penetration of
vortices and antivortices over the interface area (see Fig 2). The vortex
velocity in this case is small and it decreases the Joule term which is of
the order of $v^{2}$. Substituting functions $N_{+},N_{-}$ from the Eqs.(\ref%
{n14ab}),(\ref{n14b}),(\ref{n23}) into Eq.(\ref{n36a}) one obtains for the $%
w_{J}\simeq N_{0}\left( \nabla b_{0}\right) ^{2}\simeq I^{5/3}$ and $%
w_{A}\sim N_{0}^{2}\nabla b_{0}\simeq I^{4/3}\ $allowing to neglect in our
consideration the Joule term which is relatively small $w_{J}/w_{A}\simeq
I^{1/3}<<1$.

Here the dimensionless velocity $v$ has the form 
\begin{equation}
v=\exp \left( -U/T\right) \left\vert \frac{\partial b}{\partial x}\right\vert
\label{n37}
\end{equation}%
\bigskip while $\kappa \rightarrow t_{0}\kappa _{T}/c_{p}\left( \Delta
L\right) ^{2}$ (here $\kappa _{d}=\kappa _{T}/C_{p}$ is the diffusion
constant), $r\rightarrow t_{0}/t_{R}$ are the dimensionless effective
diffusion and relaxation coefficients correspondingly.

\subsection{Small fluctuations}

Looking for a solution of the form%
\begin{equation}
b\left( x,y,t\right) =b_{0}\left( x\right) +\psi \left( x,y,t\right) ,
\label{n46}
\end{equation}%
\begin{equation}
N\left( x,y,t\right) =N_{0}+\zeta \left( x,y,t\right) ,  \label{n47}
\end{equation}%
\begin{equation}
\Theta \left( x,y,t\right) =1+\theta \left( x,y,t\right) ,  \label{n49}
\end{equation}%
where $\psi ,$ $\zeta $ and $\theta $ are the small perturbations of the form%
\begin{equation}
\left( 
\begin{array}{c}
\psi \left( x,y,t\right) \\ 
\zeta \left( x,y,t\right) \\ 
\theta \left( x,y,t\right)%
\end{array}%
\right) =\left( 
\begin{array}{c}
\psi \left( x\right) \\ 
\zeta \left( x\right) \\ 
\theta \left( x\right)%
\end{array}%
\right) \exp \left( \lambda t+iky\right) ,  \label{n55}
\end{equation}

Here $\psi _{0},$ $\zeta _{0}$ and $\theta _{0}$ are constant amplitudes, $%
\lambda $ is the rate grow of the perturbation and $k$ is the wave vector in
the $y$ direction.

Taking into account the fluctuations of the vortex velocity%
\begin{equation}
v_{\alpha }=-\left( 1+\theta \right) \gamma _{\alpha \beta }\frac{\partial b%
}{\partial x_{\beta }}  \label{n50}
\end{equation}%
where $\theta /T_{0}$ is the change of velocity due to thermal fluctuations 
\cite{Shapiro} and $\gamma =\varphi _{0}/4\pi \eta $ (see Eq. \ref{viscosity}%
)) one obtains%
\begin{eqnarray}
v_{x} &=&v_{x}^{0}+\delta v_{x}=-\gamma _{xx}\frac{\partial b_{0}}{\partial x%
}-\gamma _{xx}\frac{\partial \psi }{\partial x}-\gamma _{xx}\frac{\partial
b_{0}}{\partial x}\theta -\gamma _{xy}\frac{\partial \psi }{\partial y}
\label{Velocity1} \\
v_{y} &=&v_{y}^{0}+\delta v_{y}=-\gamma _{yx}\frac{\partial b_{0}}{\partial x%
}-\gamma _{yx}\frac{\partial \psi }{\partial x}-\gamma _{yx}\frac{\partial
b_{0}}{\partial x}\theta -\gamma _{yy}\frac{\partial \psi }{\partial y} 
\notag
\end{eqnarray}

Here $\delta v_{x,y}$ are the deviations from the steady state vortex
velocity.

While the vortex velocity in the $x$ direction is higher then in the $y$
direction, we assume that $\left\vert v\right\vert \approx v_{x}$.

We neglect the influence of temperature fluctuations on heat capacity $C_{p}$
and relaxation coefficient $r$ because their calculations do not result in
essential effects. We also neglect in the main order the change of the
average temperature in the flux front area.

Substituting the perturbations in the form \ref{n55} into initial set of
Eqs. (\ref{n34})-(\ref{n36}), and use the stationary solution in the form
Eqs. (\ref{n24a})-(\ref{n24b}) (see Appendix I) one obtains from the Eq.(\ref%
{A10}) for the rate grow%
\begin{equation}
\lambda ^{3}+\lambda ^{2}\left( \Gamma _{1}+\Pi _{1}k^{2}\right) +\lambda
\left( \Gamma _{2}+\Pi _{2}k^{2}+\Pi _{3}k^{4}\right) +\left( \Pi
_{4}k^{2}+\Pi _{5}k^{4}\right) =0,  \label{n61}
\end{equation}%
where $\gamma _{xy}=\gamma _{yx}$ and 
\begin{eqnarray}
\Gamma _{1} &=&2\gamma _{xx}I+r-\gamma _{xx}I^{\frac{4}{3}}S_{A}
\label{n61a} \\
\Pi _{1} &=&-\gamma _{yy}I^{\frac{1}{3}}+\kappa  \notag \\
\Gamma _{2} &=&2\gamma _{xx}Ir  \notag \\
\Pi _{2} &=&2\left( \gamma _{xy}^{2}-\gamma _{xx}\gamma _{yy}\right) I^{%
\frac{4}{3}}+2\gamma _{xx}I\kappa -\gamma _{yy}I^{\frac{1}{3}}r+I^{\frac{5}{3%
}}S_{A}\left( \gamma _{xx}\gamma _{yy}-\gamma _{xy}^{2}\right)  \notag \\
\Pi _{3} &=&-\gamma _{yy}I^{\frac{1}{3}}\kappa  \notag \\
\Pi _{4} &=&2\left( \gamma _{xy}^{2}-\gamma _{xx}\gamma _{yy}\right) rI^{%
\frac{4}{3}}+\gamma _{xx}\gamma _{xy}^{2}S_{A}I^{\frac{8}{3}}  \notag \\
\Pi _{5} &=&2\left( \gamma _{xy}^{2}-\gamma _{xx}\gamma _{yy}\right) \kappa
I^{\frac{4}{3}}  \notag
\end{eqnarray}

The roots of these equations are presented in Appendix where the solution $%
\lambda _{1,2}$ are relevant.

(We consider only the nonuniform instability, hence solutions with $\func{Re}%
\lambda $ \TEXTsymbol{>}0 at $k=0$ are omitted). The onset of the nonuniform
along the front instability is determined either by the conditions $\left( 
\func{Re}\lambda _{1}=0\right) $ 
\begin{eqnarray}
\left( \func{Re}B_{1}\right) ^{2} &=&4\left( \func{Re}C_{1}\right) \left( 
\func{Re}A_{1}\right)  \label{Det} \\
\func{Re}C_{1} &<&0,\left( \func{Re}A_{1}\right) <0
\end{eqnarray}

giving the contact point at $\func{Re}\lambda _{1}-k^{2}$ plane (see figs
3-5) 
\begin{equation}
k^{2}=-\frac{\func{Re}B_{1}}{2\func{Re}A_{1}}  \label{contact}
\end{equation}%
or at $C_{1}=0$ $\left( \func{Re}B_{1}>0,\func{Re}A_{1}<0\right) $ resulting
in the Mullens-Sekerka instability.

\section{Results}

\subsection{Strong Heating.}

\bigskip

We start with a model case when the heating coefficient $S\rightarrow \infty
.$ In this case the parameters of the dispersion equation \ref{n61a} have
the form

\bigskip

\begin{eqnarray}
\Gamma _{1} &=&-\gamma _{xx}I^{\frac{4}{3}}S_{A};\Pi _{1}=-\gamma _{yy}I^{%
\frac{1}{3}}+\kappa ;\Gamma _{2}=2\gamma _{xx}Ir;\Pi _{2}=I^{\frac{5}{3}%
}S_{A}\left( \gamma _{xx}\gamma _{yy}-\gamma _{xy}^{2}\right) ;
\label{param} \\
\Pi _{3} &=&-\gamma _{yy}I^{\frac{1}{3}}\kappa ;\Pi _{4}=\gamma _{xx}\gamma
_{xy}^{2}S_{A}I^{\frac{8}{3}};\Pi _{5}=2\left( \gamma _{xy}^{2}-\gamma
_{xx}\gamma _{yy}\right) \kappa I^{\frac{4}{3}};\sqrt{\Gamma
_{1}^{2}-4\Gamma _{2}}=\gamma _{xx}I^{4/3}S_{A}  \notag
\end{eqnarray}

\bigskip while the equation for $\func{Re}\lambda _{2}$ function reads (see
Appendix II) 
\begin{equation}
\func{Re}\lambda _{2}=\frac{\gamma _{xy}^{2}S_{A}I^{\frac{5}{3}}}{4r}k^{2}-%
\frac{\gamma _{xy}^{4}S_{A}^{3}I^{\frac{11}{3}}}{64r^{3}}k^{4}
\label{Mullins}
\end{equation}%
The rapidly growing mode in this case $d^{2}\func{Re}\lambda _{2}/dk^{2}=0$
(maximum velocity of the mostly unstable mode) has the wave vector $k=2\sqrt{%
2}r/\sqrt{3}\gamma _{xy}S_{A}I$ \ and the period of the pattern along the
front (see fig.3) 
\begin{equation}
d_{y}=\sqrt{\frac{3}{2}}\frac{\pi \gamma _{xy}S_{A}I}{r}  \label{pattern}
\end{equation}%
In the case of moderate and small heating parameter $S_{A}$, the results are
strongly depends both on anisotropy and on the relation between other
parameters and can be done numerically.

\bigskip

\subsection{Moderate and weak heating.}

In this case the equation for real part of the increment of the instability
is determined by the equation (see Appendix II)

\begin{equation}
\func{Re}\lambda _{1}=\func{Re}C_{1}+k^{2}\func{Re}B_{1}+k^{4}\func{Re}A_{1};
\label{eq}
\end{equation}%
It has been solved for different anisotropy, in-plane diffusion coefficients 
$\kappa $ and relaxation coefficients $r.$The results are presented in figs.
4-6. In all of the curves at \ these figures the $\func{Re}C_{1}<0,\func{Re}%
B_{1}>0\ $and $\func{Re}A_{1}<0.$

The instabilities in all of these cases has the form of contact one rather
than the Mullens-Sekerka type. The nonuniform structure along the front
appears with the period $d_{y}=2\pi /k_{c}$ where $k_{c}$ is the contact
point of the $\func{Re}\lambda _{1}$ with $k$ \ axis.

In fig.4 the $\func{Re}\lambda _{1}$ as a function of $k^{2}$ is shown for
anisotropic superconductors with various in-plane diffusion constant $\kappa 
$. There is a critical heating parameter $S$ and critical diffusion constant
when the instability arises (curve 1), while the system becomes stable as
the diffusion constants increase (curve 2,3). The relaxation constant $r$
(the coefficient of the ballistic heat conductivity) also strongly affect
the instability condition. In fig.5 the increment $\func{Re}\lambda $ versus 
$k^{2}$ for anisotropic superconductor with different relaxation constant $r$
\ demonstrate that the instability appearing at relatively small constant $r$
(see curve 1) disappears as the relaxation parameter grow (curves 2,3).

Fig.6 demonstrates that the anisotropy essentially affects the onset of the
instability. The increment of instability $\func{Re}\lambda $ versus $k^{2}$
for different anisotropy. Curve 1 for isotropic superconductor ($\alpha
=0.9,\vartheta =\pi /4,\gamma _{xx}=\gamma _{yy}=0.545,\gamma _{xy}=0.055)$
shows the instability at heating coefficient $S_{A}=0.89$ while curve 2
demonstrates the lack of instability at heating coefficient $S_{A}=0.8.$ The
curve 3 exhibits instability for anisotropic superconductor ($\alpha
=0.1,\vartheta =\pi /4,\gamma _{xx}=\gamma _{yy}=0.55,\gamma _{xy}$ $=0.45)$
and even more small heating coefficient $S_{A}=0.1.$

\section{\protect\bigskip Conclusions}

The standing flux-antiflux interface demonstrates instability due to the
heat released by the flux-antiflux annihilation. In fact this is a well
known Kelvin-Helmholtz (KH) instability appearing when different layers of
liquid move with the opposite directed velocities \cite{KH}. In our case,
however, vortex and antivortex "liquids" are moving as it is shown in Fig.7.
The heat released by the annihilation enhances the vortex/antivortex
velocities resulting in turbulence instability at the flux-antiflux
interface. The rate grow dependence on the wave vector directed along the
front showing the instability is presented in Figs.3-6 for different heating
parameters and anisotropy of the superconducting materials. The
characteristic size of the unstable pattern is determined either by the
rapidly growing mode of the real part of the rate grow $d\lambda _{2}/dk$
(for strong heating parameter) (see fig.3) or at the contact mode for
moderate and small heating (figs.4-6).

The theory predicts stability of the flux-antiflux front for any physical
parameter of the system (see Eqs. A12 from Appendix II). The physical reason
for the instability is the result of growing temperature gradients along the
front when vortices are moving with different velocities. The velocity of
the flux flow vortices is very high and more rapid parts of the front can
break it during the time of the instability. In particular for materials
with typical parameters%
\begin{equation}
B=2000G,\eta =5\cdot 10^{-5}CGSE,v_{F}=10^{7}cm/\sec ,l=10^{-8}cm,\xi
=10^{-6}cm  \label{numbers}
\end{equation}%
where $B,\eta ,v_{F},l$ are the magnetic induction, viscosity, Fermi
velocity and mean path length of the electron correspondingly, $n_{m}=B/\phi
_{0}\simeq 10^{10}cm^{-2}$ one obtains for characteristics units of time,
space and diffusion constant $\kappa _{d}$(see Eq.\ref{n20})%
\begin{equation}
t_{0}\simeq 5\cdot 10^{-8}\sec ,\Delta L\simeq 10^{-4}cm,\kappa _{d}=\frac{%
\left( \Delta L\right) ^{2}}{t_{0}}\kappa \simeq \kappa  \label{est}
\end{equation}%
the characteristic size of the interface in the dimension units $L_{c}\simeq
\Delta L/I^{1/3}.$

For BCS superconductor where $\Delta \rightarrow 0$ the dimensional heating
parameter $S_{A}\simeq n_{m}v_{F}^{2}/T_{0}^{2}\simeq \xi ^{2}n_{m}<1$ (here 
$C_{p}\simeq mp_{F}T,$ while $\varepsilon _{F}$ and $p_{F}$ are the Fermi
energy and the momentum correspondingly. At low temperature $\ T<\Delta
_{0}\ ,$\ where $C_{p}\simeq \left( mp_{F}\Delta
_{0}^{5/2}/T_{0}^{3/2}\right) \exp \left( -\Delta _{0}/T\right) $ the
heating parameter $S_{A}$ grows dramatically $S_{A\ }\simeq \frac{%
n_{m}v_{F}^{2}}{\Delta _{0}^{2}}\left( \frac{T_{0}}{\Delta _{0}}\right)
^{1/2}\exp \left( \Delta _{0}/T\right) \gg 1.$

The heat parameter $S_{A}$ is responsible for type of the instability. In
particular at low temperatures $\left( T_{0}<<T_{c},\text{where }T_{c}\text{
is the critical temperature}\right) $ the parameter $S_{A}$ is large and the
instability develops on Mullens-Sekerka scenario (see Fig.3) typical for
dendritic instability \cite{Duran}. On the other hand, at temperatures close
to the critical, when the heating parameter $S_{A}<1$ , the instability
emerges as a periodic pattern (see figs.4-6) \cite{Langer}.

Vortices and antivortices in the unstable pattern move with velocities (see
Eq. (\ref{n33})) $u\simeq 10^{5}cm/\sec .$If the difference of the "rapid"
and "slow parts" of the front $\delta v\sim 10u$ the the flux pattern might
reach the microscopic magnitudes $L$ for very short time of the instability
development $\tau \simeq 30\mu \sec ,L\approx 3cm.$ The heat fluctuation for
this time cannot significantly relaxes because it moves along the front on
distance $\delta y\simeq \sqrt{\kappa \tau }\sim 0.01cm.$

The main results of this paper are presented in Figs 3-6 where the
increments of the instability were drown for various anisotropy parameter $%
\alpha $, heat conductivities inside and across the sample ($\kappa $ and $r$
correspondingly) and heat annihilation coefficient $S_{A}.$ We conclude that
heat released by the flux-antiflux annihilation results in instability of
the interface separating fluxons and antifluxons areas even in the case of
weak anisotropy of the superconductor.

On the other hand if the superconductor is strongly anisotropic, the
instability emerges even for weak heat. In more experimentally common case
of moderated heat and anisotropy, both these mechanisms work together
creating the instability of the flux-antiflux front.

Our major conclusion is that the anisotropy of the superconducting layered
structure alone cannot explain instability of the flux antiflux interface in
weakly anisotropic materials as $Nb$ and $MgB_{2}.$ From this point of view
without the heating, the flux-antiflux front in the $Nb$ superconductor $%
\left( \alpha \simeq 0.9\right) $ without heating should be stable for any
angle $\vartheta $ while strong heating destroys the front . If the heating
caused by the vortex-antivortex annihilation is large then the vortex
antivortex front instability should be detected in completely isotropic
superconductors like $MgB_{2}$ (Fig.6, curve 1). In fact it should be noted
that even small heating might be essentially important to cause the
instability.

This theory is appropriate in the flux flow regime. The spatial disorder
might affect the result by two different ways. It can both modify the linear
profile of the magnetic induction at the front and affect the mechanism of
the heat at the interface.

\emph{Acknowledgment }This work was supported by the Israel Academy of
Sciences (Grant 4/03-11.7).

\section{Appendix I}

Substituting Eqs.(\ref{n46})-(\ref{Velocity1}) in the initial set of Eqs. (%
\ref{n34})-(\ref{n36}), and use the stationary solution in the form Eqs. (%
\ref{n24a})-(\ref{n24b}) one obtains for perturbations%
\begin{equation}
\frac{\partial \psi }{\partial t}-N_{0}\frac{\partial \delta v_{x}}{\partial
x}-\zeta \frac{\partial v_{x}^{0}}{\partial x}-v_{x}^{0}\frac{\partial \zeta 
}{\partial x}-N_{0}\frac{\partial \delta v_{y}}{\partial y}-v_{y}^{0}\frac{%
\partial \zeta }{\partial y}=0  \label{A1}
\end{equation}%
\begin{equation}
\frac{\partial \zeta }{\partial t}-\frac{\partial b_{0}}{\partial x}\delta
v_{x}-b_{0}\frac{\partial \delta v_{x}}{\partial x}-v_{x}^{0}\frac{\partial
\psi }{\partial x}-b_{0}\frac{\partial \delta v_{y}}{\partial y}-v_{y}^{0}%
\frac{\partial \psi }{\partial y}+2\left( N_{0}\zeta -b_{0}\psi \right)
v_{x}^{0}  \label{A2}
\end{equation}%
\begin{equation}
\frac{\partial \theta }{\partial t}-\kappa \left( \frac{\partial ^{2}\theta 
}{\partial x^{2}}+\frac{\partial ^{2}\theta }{\partial y^{2}}\right)
-2S_{A}\left( N_{0}\zeta -b_{0}\psi \right) v_{x}^{0}+r\theta =0  \label{A3}
\end{equation}

Looking for the solution of these equations in the form Eq.(\ref{n55}) one
obtains near the interface where $b_{0}=-I^{2/3}x$, $N_{0}=I^{1/3}$ (see
Eqs.(\ref{n14ab})-(\ref{n14b})) the set of differential equations with
uniform coefficients:%
\begin{equation}
\left( \lambda -\gamma _{yy}I^{\frac{1}{3}}k^{2}\right) \psi \left( x\right)
+\left( \gamma _{xy}+\gamma _{yx}\right) ikI^{\frac{1}{3}}\frac{\partial
\psi \left( x\right) }{\partial x}+\gamma _{xx}I^{\frac{1}{3}}\frac{\partial
^{2}\psi \left( x\right) }{\partial x^{2}}-\gamma _{yx}ikI^{\frac{2}{3}%
}\zeta \left( x\right)  \label{A4}
\end{equation}%
\begin{equation*}
-\gamma _{xx}I^{\frac{2}{3}}\frac{\partial \zeta \left( x\right) }{\partial x%
}-\gamma _{yx}ikI\theta \left( x\right) -\gamma _{xx}I\frac{\partial \theta
\left( x\right) }{\partial x}=0
\end{equation*}%
\begin{equation}
-2\gamma _{xy}ikI^{\frac{2}{3}}\psi \left( x\right) -2\gamma _{xx}I^{\frac{2%
}{3}}\frac{\partial \psi \left( x\right) }{\partial x}+\left( \lambda
+2\gamma _{xx}I\right) \zeta \left( x\right) +\gamma _{xx}I^{\frac{4}{3}%
}\theta \left( x\right) =0  \label{A5}
\end{equation}%
\begin{equation}
\gamma _{xy}ikI^{\frac{2}{3}}S_{A}\psi \left( x\right) +\gamma _{xx}I^{\frac{%
2}{3}}S_{A}\frac{\partial \psi \left( x\right) }{\partial x}-2\gamma
_{xx}IS_{A}\zeta \left( x\right) +\left( \lambda +r+\kappa k^{2}-\gamma
_{xx}I^{\frac{4}{3}}S_{A}\right) \theta \left( x\right) -\kappa \frac{%
\partial ^{2}\theta \left( x\right) }{\partial x^{2}}=0  \label{A6}
\end{equation}

These equations should be completed by the boundary conditions

\begin{equation}
\bigskip \left( 
\begin{array}{c}
\psi \\ 
\zeta \\ 
\theta%
\end{array}%
\right) _{x=-x_{c}/2,x_{c}/2}=0  \label{boundary}
\end{equation}

The functions $\psi \left( x\right) ,$ $\zeta \left( x\right) $ and $\theta
\left( x\right) $ should be symmetrical and localized at the flux-antiflux
interface where $x<$ $x_{c}$ while $x_{c}\gg 1$ is the cutoff where these
functions go to zero (see Fig.2 and Eq. (\ref{interface})).

Looking for solution of Eqs. (\ref{A1})-(\ref{A6}) in the form

\begin{equation}
\left( 
\begin{array}{c}
\psi \\ 
\zeta \\ 
\theta%
\end{array}%
\right) =\left( 
\begin{array}{c}
\left( A_{n}\sin p_{n}x+B_{n}\cos p_{n}x\right) \\ 
\left( C_{n}\sin p_{n}x+D_{n}\cos p_{n}x\right) \\ 
\left( E_{n}\sin p_{n}x+F_{n}\cos p_{n}x\right)%
\end{array}%
\right) ;  \label{matrix}
\end{equation}%
$\ $one obtains equations for $A_{n},B_{n},C_{n},D_{n},E_{n},F_{n}$
coefficients%
\begin{equation}
\widehat{\Lambda }\left( 
\begin{array}{c}
A_{n} \\ 
B_{n} \\ 
C_{n} \\ 
D_{n} \\ 
E_{n} \\ 
F_{n}%
\end{array}%
\right) =0  \label{Equt}
\end{equation}

where matrix $\widehat{\Lambda }\ $reads%
\begin{equation}
\left( 
\begin{array}{cccccc}
\begin{array}{c}
\lambda -\gamma _{yy}I^{\frac{1}{3}}k^{2}- \\ 
-\gamma _{xx}I^{\frac{1}{3}}p_{n}^{2}%
\end{array}
& -2\gamma _{xy}ip_{n}kI^{\frac{1}{3}} & -\gamma _{yx}ikI^{\frac{2}{3}} & 
p_{n}\gamma _{xx}I^{\frac{2}{3}} & -\gamma _{yx}ikI & p_{n}\gamma _{xx}I \\ 
p_{n}2\gamma _{xy}ikI^{\frac{1}{3}} & 
\begin{array}{c}
\lambda -\gamma _{yy}I^{\frac{1}{3}}k^{2}- \\ 
-\gamma _{xx}I^{\frac{1}{3}}p_{n}^{2}%
\end{array}
& -\gamma _{xx}I^{\frac{2}{3}}p_{n} & -\gamma _{yx}ikI^{\frac{2}{3}} & 
-\gamma _{xx}Ip_{n} & -\gamma _{yx}ikI \\ 
-2\gamma _{xy}ikI^{\frac{2}{3}} & 2p_{n}\gamma _{xx}I^{\frac{2}{3}} & 
\lambda +2\gamma _{xx}I & 0 & \gamma _{xx}I^{\frac{4}{3}} & 0 \\ 
-2p_{n}\gamma _{xx}I^{\frac{2}{3}} & -2\gamma _{xy}ikI^{\frac{2}{3}} & 0 & 
\lambda +2\gamma _{xx}I & 0 & \gamma _{xx}I^{\frac{4}{3}} \\ 
\gamma _{xy}ikI^{\frac{2}{3}}S_{A} & -\gamma _{xx}I^{\frac{2}{3}}SP_{n} & 
-2\gamma _{xx}IS_{A} & 0 & 
\begin{array}{c}
\lambda +r+\kappa k^{2}- \\ 
-\gamma _{xx}I^{\frac{4}{3}}S_{A}+\kappa p_{n}^{2}%
\end{array}
&  \\ 
\gamma _{xx}I^{\frac{2}{3}}S_{A}P_{n} & \gamma _{xy}ikI^{\frac{2}{3}}S_{A} & 
0 & -2\gamma _{xx}IS_{A} & 0 & 
\begin{array}{c}
\lambda +r+\kappa k^{2}- \\ 
-\gamma _{xx}I^{\frac{4}{3}}S_{A}+\kappa p_{n}^{2}%
\end{array}%
\end{array}%
\right)  \label{A7}
\end{equation}

where $p_{n}=\left( 2n+1\right) \pi /x_{c}$ is obtained from the boundary
condition (\ref{boundary}). The dangerous harmonic with $n=0$ is responsible
for instability. In this case $p_{0}\sim x_{c}^{-1}\sim I^{1/3}\rightarrow 0$
for small slope of the magnetic induction at the interface $\left( I\ll
1\right) $ this matrix can be simplified$:$%
\begin{equation}
\left\vert 
\begin{array}{cccccc}
\lambda -\gamma _{yy}I^{\frac{1}{3}}k^{2} & 0 & -\gamma _{yx}ikI^{\frac{2}{3}%
} & 0 & -\gamma _{yx}ikI & 0 \\ 
0 & \lambda -\gamma _{yy}I^{\frac{1}{3}}k^{2} & 0 & -\gamma _{yx}ikI^{\frac{2%
}{3}} & 0 & -\gamma _{yx}ikI \\ 
-2\gamma _{xy}ikI^{\frac{2}{3}} & 0 & \lambda +2\gamma _{xx}I & 0 & \gamma
_{xx}I^{\frac{4}{3}} & 0 \\ 
0 & -2\gamma _{xy}ikI^{\frac{2}{3}} & 0 & \lambda +2\gamma _{xx}I & 0 & 
\gamma _{xx}I^{\frac{4}{3}} \\ 
\gamma _{xy}ikI^{\frac{2}{3}}S_{A} & 0 & -2\gamma _{xx}IS_{A} & 0 & \lambda
+r+\kappa k^{2}-\gamma _{xx}I^{\frac{4}{3}}S_{A} & 0 \\ 
0 & \gamma _{xy}ikI^{\frac{2}{3}}S_{A} & 0 & -2\gamma _{xx}IS_{A} & 0 & 
\lambda +r+\kappa k^{2}-\gamma _{xx}I^{\frac{4}{3}}S_{A}%
\end{array}%
\right\vert =0  \label{A9}
\end{equation}

\bigskip and can be represented as generated Jordan matrix.

\begin{equation}
\left\vert 
\begin{array}{cc}
\widehat{K} & 0 \\ 
0 & \widehat{K}%
\end{array}%
\right\vert =0  \label{A9a}
\end{equation}%
where $K$ is the $3\times 3$ matrix

\begin{equation}
\widehat{K}=\left\vert 
\begin{array}{ccc}
\left( \lambda -\gamma _{yy}I^{\frac{1}{3}}k^{2}\right) & \left( -\gamma
_{yx}ikI^{\frac{2}{3}}\right) & \left( -\gamma _{yx}ikI\right) \\ 
\left( -2\gamma _{xy}ikI^{\frac{2}{3}}\right) & \left( \lambda +2\gamma
_{xx}I\right) & \left( \gamma _{xx}I^{\frac{4}{3}}\right) \\ 
\left( \gamma _{xy}ikI^{\frac{2}{3}}S_{A}\right) & \left( -2\gamma
_{xx}IS_{A}\right) & \left( \lambda +r+\kappa k^{2}-\gamma _{xx}I^{\frac{4}{3%
}}S_{A}\right)%
\end{array}%
\right\vert  \label{A10}
\end{equation}

\bigskip giving the equation for in the form \bigskip 
\begin{equation}
\lambda ^{3}+\lambda ^{2}\left( \Gamma _{1}+\Pi _{1}k^{2}\right) +\lambda
\left( \Gamma _{2}+\Pi _{2}k^{2}+\Pi _{3}k^{4}\right) +\left( \Pi
_{4}k^{2}+\Pi _{5}k^{4}\right) =0,  \label{A11}
\end{equation}%
\begin{eqnarray*}
\Gamma _{1} &=&2\gamma _{xx}I+r-\gamma _{xx}I^{\frac{4}{3}}S_{A} \\
\Pi _{1} &=&-\gamma _{yy}I^{\frac{1}{3}}+\kappa \\
\Gamma _{2} &=&2\gamma _{xx}Ir \\
\Pi _{2} &=&2\left( \gamma _{xy}^{2}-\gamma _{xx}\gamma _{yy}\right) I^{%
\frac{4}{3}}+2\gamma _{xx}I\kappa -\gamma _{yy}I^{\frac{1}{3}}r+I^{\frac{5}{3%
}}S_{A}\left( \gamma _{xx}\gamma _{yy}-\gamma _{xy}^{2}\right) \\
\Pi _{3} &=&-\gamma _{yy}I^{\frac{1}{3}}\kappa \\
\Pi _{4} &=&2\left( \gamma _{xy}^{2}-\gamma _{xx}\gamma _{yy}\right) rI^{%
\frac{4}{3}}+\gamma _{xx}\gamma _{xy}^{2}S_{A}I^{\frac{8}{3}} \\
\Pi _{5} &=&2\left( \gamma _{xy}^{2}-\gamma _{xx}\gamma _{yy}\right) \kappa
I^{\frac{4}{3}}
\end{eqnarray*}

\bigskip

\section{\protect\bigskip Appendix II}

\bigskip The solutions of the Eq. (\ref{A11}) at small $k\rightarrow 0$
read: 
\begin{eqnarray}
\lambda _{0} &=&-\frac{\Pi _{4}}{\Gamma _{2}}k^{2}+A_{0}k^{4}  \label{A12} \\
A_{0} &=&\frac{\Pi _{2}\Pi _{4}}{\Gamma _{2}^{2}}-\frac{\Gamma _{1}\Pi
_{4}^{2}}{\Gamma _{2}^{3}}-\frac{\Pi _{5}}{\Gamma _{2}}  \notag
\end{eqnarray}

\bigskip

\bigskip 
\begin{eqnarray}
\lambda _{1} &=&C_{1}+B_{1}k^{2}+A_{1}k^{4};  \label{A13} \\
C_{1} &=&\frac{1}{2}\left[ -\Gamma _{1}+\sqrt{\Gamma _{1}^{2}-4\Gamma _{2}}%
\right] ; \\
B_{1} &=&\frac{\left[ \Gamma _{1}-\sqrt{\Gamma _{1}^{2}-4\Gamma _{2}}\right] %
\left[ \Gamma _{1}\Pi _{1}-\Pi _{2}\right] +2\left[ \Pi _{4}-\Gamma _{2}\Pi
_{1}\right] }{\Gamma _{1}\left[ \sqrt{\Gamma _{1}^{2}-4\Gamma _{2}}-\Gamma
_{1}\right] +4\Gamma _{2}} \\
A_{1} &=&\frac{B_{1}^{2}\left[ 3\sqrt{\Gamma _{1}^{2}-4\Gamma _{2}}-\Gamma
_{1}\right] +\left[ \sqrt{\Gamma _{1}^{2}-4\Gamma _{2}}-\Gamma _{1}\right] %
\left[ 2B_{1}\Pi _{1}+\Pi _{3}\right] +2\left[ \Pi _{5}+B_{1}\Pi _{1}\right] 
}{4\Gamma _{2}+\Gamma _{1}\left[ \sqrt{\Gamma _{1}^{2}-4\Gamma _{2}}-\Gamma
_{1}\right] }
\end{eqnarray}%
and 
\begin{eqnarray}
\lambda _{2} &=&C_{2}+B_{2}k^{2}+A_{2}k^{4};  \label{A14} \\
C_{2} &=&\frac{1}{2}\left[ -\Gamma _{1}-\sqrt{\Gamma _{1}^{2}-4\Gamma _{2}}%
\right] ;  \notag \\
B_{2} &=&\frac{\left[ \Gamma _{1}+\sqrt{\Gamma _{1}^{2}-4\Gamma _{2}}\right] %
\left[ \Pi _{2}-\Gamma _{1}\Pi _{1}\right] +2\left[ \Gamma _{2}\Pi _{1}-\Pi
_{4}\right] }{\Gamma _{1}\left[ \sqrt{\Gamma _{1}^{2}-4\Gamma _{2}}+\Gamma
_{1}\right] -4\Gamma _{2}}; \\
A_{2} &=&\frac{B_{2}^{2}\left[ 3\sqrt{\Gamma _{1}^{2}-4\Gamma _{2}}+\Gamma
_{1}\right] +\left[ \sqrt{\Gamma _{1}^{2}-4\Gamma _{2}}+\Gamma _{1}\right] %
\left[ 2B_{2}\Pi _{1}+\Pi _{3}\right] -2\left[ \Pi _{5}+B_{2}\Pi _{1}\right] 
}{\Gamma _{1}\left[ \sqrt{\Gamma _{1}^{2}-4\Gamma _{2}}+\Gamma _{1}\right]
-4\Gamma _{2}}
\end{eqnarray}%
\newpage

Figure Captions

Fig.1\qquad Geometry of the problem. $\vartheta $ is the angle between the $%
x $ axis and the $ab$ plain of the layered structure and $c$ axis is
perpendicular to the layers of the superconductor.

Fig.2 Structure of the vortex-antivortex interface.

Fig.3 Mullins-Sekerka instability for super large heat. The increment $\func{%
Re}\lambda $ versus $k^{2}$ ($k$ is the wave vector along the flux-antiflux
front).

Fig.4\qquad The increment $\func{Re}\lambda $ versus $k^{2}$ for isotropic
superconductors $\left( \alpha =0.9,\vartheta =\pi /4\right) $with different
in-plane diffusion constant $\kappa .$Curves 1,2,3 correspond to $\kappa
=0.1;0.5;1$ respectively. Here $\gamma _{xx}=\gamma _{yy}=0.545,\gamma
_{xy}=0.055,I=0.5,r=0.0148,S_{A}=0.89.$ Instability disappears as the
diffusion constant grows.

Fig.5\qquad The increment $\func{Re}\lambda $ versus $k^{2}$ for anisotropic
superconductor with different relaxation constant $r$ . Curves 1,2,3
correspond to parameters $r=0.049;0.13;2.66$ respectively. The instability
disappears as the relaxation parameter grows. Here $\gamma _{xx}=\gamma
_{yy}=0.545,\gamma _{xy}=0.055,\kappa =0.1,I=0.5,S_{A}=0.89.$

Fig.6\qquad The increment of instability $\func{Re}\lambda $ versus $k^{2}$
for different anisotropy. Curve 1 for isotropic superconductor ($\alpha
=0.9,\vartheta =\pi /4,\gamma _{xx}=\gamma _{yy}=0.545,\gamma _{xy}=0.055)$
shows the instability at heating coefficient $S_{A}=0.89$ while curve 2
demonstrates the lack of instability at heating coefficient $S_{A}=0.8.$ The
curve 3 exhibits instability for anisotropic superconductor ($\alpha
=0.1,\vartheta =\pi /4,\gamma _{xx}=\gamma _{yy}=0.55,\gamma _{xy}$ $=0.45)$
and even more small heating coefficient $S_{A}=0.1.$(Here $r=0.0148,\kappa
=0.1,I=0.5).$

Fig.7 \ \ \ \ \ Qualitative picture of the Kelvin-Helmholtz instability at
the vortex-antivortex interface. 


\newpage

\begin{thebibliography}{99}
\bibitem{Johansen} Johansen T H, Bazilevich M, Bratsberg H, Hauglin H and
Lafyatis G 1996 \textit{High Temperature Superconductors: Synthesis,
Processing, and Large-Scale Applications} edited by U Balachandran P J
McGinn and J S Abell (The Minerals, Metals \& Materials Society, New York) p
203.

\bibitem{Vlasko-Vlascov} Vlasko-Vlasov V K \textit{et al }1994 \textit{%
Physica} C \textbf{222} 361.

\bibitem{Koblischka} Koblischka M R Johansen T H Bazilevich M Hauglin H
Bratsberg H and Shapiro B Ya 1998 \textit{Europhys. Lett.} \textbf{41} 419.

\bibitem{Duran} Duran C A Gammel P L Miller R E and Bishop D J 1995 \textit{%
Phys. Rev.} B \textbf{52} 75.

\bibitem{Leiderer} Leiderer P Boneberg J Br\"{u}ll P Bujok V and Herminghaus
S 1993 \textit{Phys. Rev. Lett.} \textbf{71} 2646.

\bibitem{Yeshurun} Barness D Sinvani M Shaulov A Tamegai T and Yeshurun Y
2008 \textit{Phys. Rev.} B \textbf{77} 094514.

\bibitem{Johansen02} Johansen T H Baziljevich M Shantsev D V Goa P E
Galperin Y M Kang W N Kim H J Choi E M Kim M S and Lee S I 2002 \textit{%
Europhys. Lett.} \textbf{59} 599.

\bibitem{Barkov} Barkov F L Shantsev D V Johansen T H Goa P E Kang W N Kim H
J Choi E M and Lee S I 2003 \textit{Phys. Rev.} B \textbf{67} 064513.

\bibitem{Marchevsky} Marchevsky M Gurevich L A Kes P H and Aarts J 1995 
\textit{Phys. Rev. Lett.} \textbf{75} 2400.

\bibitem{Mullins} Mullins W W and Sekerka R F 1963 \textit{J. Appl. Phys.} 
\textbf{34} 323.

\bibitem{Mints} Mints G R and Rachmanov A L 1981 \textit{Rev. Mod. Phys.} 
\textbf{53} 551.

\bibitem{Aranson01} Aranson I Gurevich A and Vinokur V 2001 \textit{Phys.
Rev. Lett.} \textbf{87} 067003.

\bibitem{Rakhmanov04} Rakhmanov A L Shantsev D V Galperin Y M and Johansen T
H 2004 \textit{Phys. Rev.} B \textbf{70} 224502.

\bibitem{Aranson} Aranson I Shapiro B Ya and Vinokur V 1996 \textit{Phys.
Rev. Lett.} \textbf{76} 142.

\bibitem{Frello} Frello T Baziljevich M Johansen T H Andersen N H Wolf Th
and Koblischka M M R 1999 \textit{Phys. Rev.} B \textbf{59} R6639.

\bibitem{Indebom} Indebom M V Kronm\"{u}ller H Kes P and Menovsky A A 1993 
\textit{Physica} C \textbf{209} 259.

\bibitem{Fisher} Fisher L M Goa P E Baziljevich M Johansen T H Rakhmanov A L
and Yampol'skii V A 2001 \textit{Phys. Rev. Lett.} \textbf{87} 247005.

\bibitem{Rakhmanov} Rakhmanov A L Fisher L M Levchenko A A Yampolskii V A
Baziljevich M and Johansen T H 2002 \textit{JETP Lett.} \textbf{76} 291.

\bibitem{Van Saarloos} Baggio C Howard M and van Saarloos W 2004 \textit{%
Phys. Rev.} E \textbf{70} 026209.

\bibitem{Bryksin} Bryksin V V and Dorogovtsev S N 1993 \textit{Zh. Eksp.
Teor. Fiz.} \textbf{104} 3735 [\textit{Sov. Phys. JETP} \textbf{77} 791].

\bibitem{Landau10} Landau L D and Lifshits E M 1991 \textit{Physical Kinetics%
} Pergamon Oxford.

\bibitem{Bakhanova} Bakhanova E S Genkin V M Kalyagin M A Konkin S N and
Churin S A 1991 \textit{Zh. Eksp. Teor. Fiz.} \textbf{100} 1919 [\textit{%
Sov. Phys. JETP} \textbf{73} 1061].

\bibitem{Shapiro} Bass F Shapiro B Ya and Shvartser M 1998 \textit{Phys.
Rev. Lett.} \textbf{80} 2441; Bass F Shapiro B Ya Shapiro I and Shvartser M
1998 \textit{Phys. Rev. B} \textbf{58} 2878.

\bibitem{Langer} Langer J S 1980 \textit{Rev. Mod. Phys.} \textbf{52} 1.

\bibitem{KH} Drazin, P. G. and W. H. Reid, 1981: \textit{Hydrodynamic
Stability}. Cambridge Univ. Press, London.

\newpage 
\end{thebibliography}
\end{document}